\documentclass[%
 reprint,
nofootinbib,
 amsmath,amssymb,
 aps,
floatfix,
]{revtex4-2}

\usepackage{graphicx}
\usepackage{svg}
\graphicspath{{figures/}}

\usepackage{dcolumn}
\usepackage{bm}
\usepackage{hyperref}
\usepackage[capitalize]{cleveref}
\usepackage{physics}

\usepackage{xcolor}
\usepackage{lettrine}

\begin{document}

\preprint{APS/123-QED}

\title{Quantum Super-Resolution with Balanced Homodyne Detection in Low-Earth-Orbit}

\author{Ronakraj K. Gosalia}
 \email{r.gosalia@unsw.edu.au}
 \author{Robert Malaney}
\affiliation{
 University of New South Wales, Sydney, NSW 2052, Australia.
}
\author{Ryan Aguinaldo}
\author{Jonathan Green}
\affiliation{
 Northrop Grumman Corporation, San Diego, CA 92128, USA.
}


\begin{abstract}
Quantum super-resolution involves resolving two sources below the Rayleigh limit using quantum optics. Such a technique would allow high-precision inter-satellite positioning and tracking on communication and navigation constellations. Due to the size, weight and power constraints typical of low-earth-orbit (LEO) satellites, a simple solution is often preferred. Here, we show that a balanced homodyne detection (BHD) setup using a shaped single-mode local oscillator can achieve super-resolution despite typical photonic losses. We further analyze the impact of a fluctuating and fixed centroid misalignment due to satellite pointing issues, and find that fixed misalignment is comparatively more detrimental to the performance of a BHD setup. Thus, our study provides a practical assessment of BHD to achieve super-resolution on a modern LEO satellite platform. Finally, we discuss how our analysis can be extended to stellar sources for astronomical applications. 
\end{abstract}

\maketitle


\section{\label{sec:intro}Introduction}
\lettrine{O}{}ver the last few decades, considerable attention has been given to finding the ultimate limit of resolution in an optical system. In many works, the Rayleigh criterion was simply taken to be this limit\footnote{In practice this criterion is of limited pragmatic value as the noise in the observations must be taken into account in any realizable resolution limit (e.g. \cite{limitray}). Nonetheless, we use the traditional ``noiseless" Rayleigh limit as our comparison in this work.}. The criterion states that to resolve two sources of light, the first diffraction minimum of one source must coincide with the maximum of the other. In our context, $d_{min}\geq \lambda\ell/(4r)= d_{rayleigh}$, where $2d_{min}$ is the minimum resolvable transverse separation between two sources. Here, $\lambda$ is the laser central wavelength, $\ell$ is the propagation distance between source and receiver and $r$ denotes the radius of the receiver aperture.

However, quantum optics has provided a pathway to overcome this Rayleigh criterion. Efforts using quantum optics to achieve super-resolution can yield a resolution that is ultimately limited not by diffraction, but by the quantum fluctuations of light \cite{Kolobov2000,Tamburini2006,Tsang2019}. By modeling light as a quantum system, the analysis of the resolution limit accounts for both diffraction and photon shot noise \cite{Tsang2016}. In other words, a quantum-approach can enable the development of optical systems that reach the fundamental limits of resolution \cite{Delaubert2008,Perez-Delgado2012,Tsang2019,Lupo2020}. Herein we use a quantum approach to resolve two sources that are coherent with respect to each other in the regime of super-resolution which we define here to be any system setup where $d_{min}<d_{rayleigh}$. 

Recent efforts have focused on the problem of achieving super-resolution with sources that are incoherent with respect to each other via quantum estimation theory \cite{Tsang2016,Tsang2018,Tsang2019}. Particularly, techniques such as spatial-mode demultiplexing (SPADE) \cite{Tsang2018,Len2020,Boucher2020a} and super-resolved position localization by inversion of coherence along an edge (SPLICE) \cite{Tham2017,Bonsma-fisher2019,Qi2022} have shown promising pathways for fields including astronomy \cite{Tsang2019,Tao2020, Zanforlin2022} and microscopy \cite{Ram2006,Deschout2014,Chao2016}. However, a technique that has received relatively less attention is balanced homodyne detection (BHD) with a shaped local oscillator (LO) --- henceforth, simply referred to as BHD. 

It is known that BHD, like SPADE, is an optimal method for resolving two sources when $d_{min}<d_{rayleigh}$ \cite{Hsu2004,Delaubert2006,Delaubert2008,Pinel2012b,Yang2017,Brossel2018}. However, for super-resolution imaging of incoherent (thermal) sources, such as stellar light, it was found that an average of more than two photons per mode per measurement time is required --- which is an extremely challenging feat \cite{Yang2017a,Tsang2019}. Fortunately, the same issue is not expected over inter-satellite laser links due to shorter transmit-receiver range and more control over the source light. Particularly, in our scenario of interest, low-Earth-orbit (LEO) satellite networks, BHD is already gaining momentum as a technique for next-generation communications \cite{Kish2020}, clock synchronization \cite{Lamine2008,Gosalia2022,Gosalia2023} and laser ranging \cite{Tian2019}. The analysis presented here shows that BHD may also enable high-precision positioning and navigation on LEO satellite constellations.
 
It is yet to be determined whether BHD can sustain optimal super-resolution performance whilst on-board a real LEO satellite --- the main issue we investigate here. 
We analytically study the performance of BHD using a $\text{HG}_{1,0}$ spatial-mode LO. Our contribution is to answer whether there are any realistic transmitter-receiver ranges (and noise conditions) under which the BHD setup can achieve super-resolution in LEO. We answer in the affirmative, detailing scenarios where super-resolution is achievable despite photonic losses due to diffraction, detection inefficiencies inherent to detectors, and noise from satellite pointing issues. To make progress we will focus on two application spaces. One is the separation of one (two) coherent source(s) in space from some pre-defined axis, and the other is the resolution of incoherent (thermal) sources in the astronomical context. 

The remainder of this paper is  as follows. In \cref{sec:system}, we detail the main system model 
and formalize the excitation of higher-order spatial modes due to the separation distance, $2d$. The FI of the BHD setup is presented in \cref{sec:fisher}, and the impact of photonic loss in a typical LEO environment is investigated in \cref{sec:snr}. In \cref{sec:centroid}, fluctuating (\cref{sec:centroid-fluctuating}) and fixed (\cref{sec:centroid-fixed}) centroid misalignment issues are studied as well as their impact on super-resolution. 
Finally, 
\cref{sec:conc} concludes the work.

\section{\label{sec:system}Analysis}
\begin{figure}
    \centering
    \includegraphics[width=\linewidth]{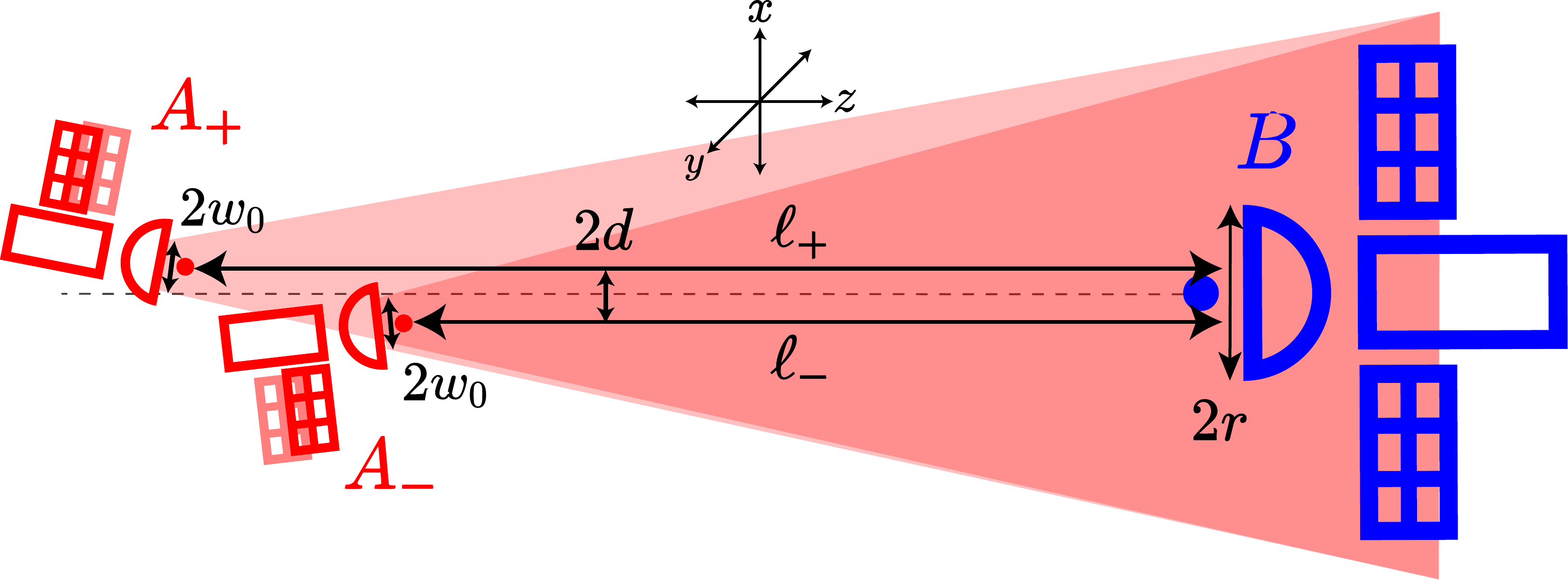}
    \caption{\label{fig:satellite-layout}Overall system diagram showing two sources, $A_{\pm}$, each transmitting laser beams (that are temporally coherent with respect to each other) to a remote receiver, $B$, from propagation distances $\ell_{\pm}$, respectively. The sources are separated by $2d$ and $B$ uses BHD with a $\text{HG}_{1,0}$ shaped LO to estimate $d$ when $d<w_0$ with the aim to reach super-resolution, i.e. $d_{min}<d_{rayleigh}$, over realistic inter-satellite ranges in LEO.}
\end{figure}

Consider two identical sources $A_+$ and $A_-$ (denoted together as $A_{\pm}$) that have a small separation represented by $2\vec{d}$ along the transverse $xy$-plane. Both sources emit identical laser beams to a remote receiver, $B$, from propagation distances $\ell_{\pm}$, respectively, along the $z$-axis. 
We want to estimate $\lvert\lvert\vec{d}\rvert\rvert$ (length of $\vec{d}$, henceforth simply denoted as $d$) when $d<w_0$ using a BHD setup with a shaped LO. Further, we want to ascertain practical scenarios in which super-resolution (i.e. $d_{min}<d_{rayleigh}$) is achievable. See \cref{fig:satellite-layout} for the system diagram. Note, although we consider two sources, the case of just one source is also readily covered by this analysis via some simple adjustments.

The key assumptions made throughout are as follows: $A_{\pm}$ transmit temporally coherent Gaussian beams with spatial profile in the $\text{HG}_{0,0}$ mode\footnote{Also denoted as the transverse electromagnetic modes $\text{TEM}_{nm}$.}; the separation distance between $A_{\pm}$, $2d$, is entirely along the $x$-axis on the $xy$ transverse plane; $A_{\pm}$ and $B$ do not undergo relative motion during measurement; all three satellites are temporally synchronized and have lasers with the same wavelength $\lambda$; and the centroid angle (between $A_{\pm}$ as viewed from $B$) is known \emph{a priori} at $B$\footnote{From \cref{sec:centroid} onward we remove this centroid assumption.}.

\subsection{\label{sec:modes}Excitation of higher-order spatial modes}
The transverse profile of a beam can be described by a set of orthogonal spatial modes $\{u_{n,m}\}$ that are solutions to the wave equation \cite{Kogelnik1966a}. If we consider a beam in the fundamental mode $u_{0,0}$, it can be represented by the positive electric-field operator
\begin{align}
    \label{eq:e00}
    \hat{E}_{0,0}(x,y,z) = \mathcal{E}u_{0,0}(x,y,z)\hat{a}_0,
\end{align}
where 
the normalization coefficient $\mathcal{E}=\sqrt{\hbar\pi c/(\epsilon_0 \lambda V)}$ with the reduced Planck constant $\hbar$, the laser central wavelength $\lambda$, the permittivity of free space $\epsilon_0$, the measurement volume $V$ and $\hat{a}_{0,0}$ ($\hat{a}^\dag_{0,0}$) are the annihilation (creation) operator of mode $u_{0,0}$. 

We can specify $\{u_{n,m}\}$ to be the HG mode basis, with $u_{n,m}(x,y,z)=u_n(x,z)u_m(y,z)$ corresponding to $\text{HG}_{n,m}$, with wave number $k=2\pi/\lambda$, and \cite{Kogelnik1966a}
\begin{align}
    \label{eq:u_n}
    &\nonumber u_{n}(x,z) = \left(\frac{2}{\pi}\right)^{1/4}\left(\frac{{\rm e}^{i(2n+1)\Psi(z)}}{2^n n!w(z)}\right)^{1/2}\times \\
    &H_{n}\left(\frac{\sqrt{2}x}{w(z)}\right)\exp\left(-x^2\left(\frac{1}{w^2(z)}+\frac{ik}{2R_c(z)}\right)\right).
\end{align}
Here, the Guoy phase $\Psi(z)=\arctan(z/z_R)$, the beam width $w(z)=w_0(1+(z/z_R)^2)^{1/2}$, the Rayleigh range $z_R~=~\pi w_0^2/\lambda$, the beam waist $w_0$, the speed of light in vacuum $c$, the physicist's Hermite polynomial $H_{n}(\cdot)$ of order $n$, and the radius of curvature $R_c(z)=z(1+(z_R/z)^2)$. Similarly, $u_m(y,z)$ can be found by substituting $x$ for $y$ in \cref{eq:u_n}.

A non-zero spatial displacement, $\vec{d}$, of a beam initially in the $\text{HG}_{0,0}$ mode will proportionally excite higher-order spatial modes. For instance, along the $x$-axis, a displacement $d_x$ on the mode $u_{n}(x,z)$ becomes \cite{Delaubert2006}
\begin{align}
    \label{eq:u_0_dx}
    \nonumber u_{0}\left(x+d_x,\ell_{\pm}\right) &= u_{0}(x,\ell_{\pm})+\sum_{j=1}^{\infty} \left(d_x\right)^j\frac{\partial^j}{\partial x^j}u_{0}(x,\ell_{\pm}) \\
    &\approx~u_0(x,\ell_{\pm}) - \frac{d_x}{w_0} u_1(x,\ell_{\pm}),
\end{align}
where the first-order truncation is appropriate only when it can be assumed that $d_x< w_0$, as is the case in this work. Here, $d_x=d\cos(\theta_d)$, $\theta_d$ is the angle between $\vec{d}$ and the $xy$-axes on the transverse plane. Note, we can also find $u_{0}(y+d_y,\ell_{\pm})$ by substituting in $d_y=d\sin(\theta_d)$ and $y$ in \cref{eq:u_0_dx}. 

The combined field description from both sources $A_{\pm}$ at $B$ is given by
\begin{align}
    \label{eq:e-in}
    \nonumber
    \hat{E}_{in}(x,y) = \frac{1}{\sqrt{2}}\left(\hat{E}_{+}(x,y,\ell_{+})+\hat{E}_{-}(x,y,\ell_{-})\right), ~\text{where} \\
    \nonumber \hat{E}_{\pm}(x,y,\ell_{\pm}) \approx\hat{E}_{0,0}(x,y,\ell_{\pm}) \mp \frac{d}{w_0}\cos(\theta_d) \hat{E}_{1,0}(x,y,\ell_{\pm})\mp\\
    \frac{d}{w_0}\sin(\theta_d) \hat{E}_{0,1}(x,y,\ell_{\pm}) + \frac{d^2}{2w_0^2}\sin(2\theta_d)\hat{E}_{1,1}(x,y,\ell_{\pm}).
\end{align}
Since our parameter of interest is $d$, it suffices to measure the intensities of the $\text{HG}_{1,0}$, $\text{HG}_{0,1}$ and $\text{HG}_{1,1}$ modes of $\hat{E}_{in}$. Going forward we make some simplifications to assist our analysis. We begin by setting $\theta_d=0$ assuming that $\vec{d}$ is entirely along the $x$-axis. This simplification allows us to focus on the single $\text{HG}_{1,0}$ mode, and is not dissimilar to 
several prior works 
\cite{Sun2014,Brossel2018,Yang2017a,Pinel2012b,Delaubert2006,Hsu2004}. 

In terms of state formalism, the incoming coherent states will can be represented by the tensor product
\begin{align}
    \nonumber
    \label{eq:coherent-sources}
    \ket{\alpha_{in}} &= \ket{\alpha_+}\otimes\ket{\alpha_-},~\text{where}\\
    \ket{\alpha_{\pm}} &\approx \ket{\alpha_\pm}_{0,0}\otimes\ket{\mp\frac{d}{w_0}\alpha_{\pm}}_{1,0}.
\end{align}
Here, $\alpha_{in}$ is the complex amplitude of the combined coherent state defined as $\alpha_{in}:=\sqrt{N_{in}}{\rm e}^{i\Phi_{in}}$, with $N_{in}$ the total number of photons from both sources and $\Phi_{in}$ the phase relative to the LO at $B$. Similarly, $\alpha_{\pm}:=\sqrt{N_{\pm}}{\rm e}^{i\Phi_{\pm}}$ are the complex amplitudes of each source, with $N_{\pm}$ denoting photon numbers and $\Phi_{\pm}$ denoting phase of each source relative to $B$'s LO. 

Finally, we note that in a practical implementation, several higher-order spatial modes will need to be simultaneously measured using a multimode LO (e.g., \cite{Hsu2004}). The analysis presented here can be extended to the general case with some effort. We also note that for large separation distances, $d> w_0$, conventional direct imaging suffices to optimally estimate $d$ \cite{Tsang2016}. In practice on LEO satellites, we therefore recommend a hybrid approach with both coarse estimation using direct imaging, and fine estimation using a BHD setup. Although, the focus of this study remains on the latter aspect.

\subsection{\label{sec:homodyne}Balanced homodyne detection}
We first conduct a Fisher information (FI) analysis to find the maximum amount of information that can be obtained about $d$ using BHD and to derive the standard quantum limit (SQL), which is the limit on the resolution of $d$ dictated by the quantum fluctuations of light. Secondly, we conduct a signal-to-noise ratio (SNR) analysis to obtain a more realistic bound on $d$ as guided by the photonic loss due to diffraction and detector inefficiency. By comparing the FI result with the SNR analysis, we can quantify the amount of loss that is tolerable for achieving super-resolution in practical LEO satellite configurations.

\subsubsection{\label{sec:fisher}Fisher information}
The quantum Cram\'er-Rao bound (q-CRB) is the fundamental limit of the precision with which a parameter can be estimated depending only on the quantum state of the probe, and optimized over all possible measurement schemes \cite{Liu2020}. Several prior studies have shown that BHD can be used as a measurement scheme to reach the q-CRB of separation distance estimation, and is therefore optimal \cite{Pinel2012b,Sun2014,Li2023}. In this work, we take this result for granted and extend the FI and classical Cram\'er-Rao bound (CRB) of BHD to consider the relative phase between two coherent sources during the measurement.

The BHD setup will measure a quadrature of the incoming field in the same mode as the LO field. We specify a LO shaped in the $\text{HG}_{1,0}$ mode since this is the excited mode of the combined light from $A_{\pm}$\footnote{In other scenarios where the source fields are in higher $\text{HG}_{n,m}$ modes see \cite{Sun2014,Xia2023}.}. As output, the $\hat{X}$ quadrature in the $\text{HG}_{1,0}$ mode of $\hat{E}_{in}$ will be sampled (denoted $\hat{X}_{in_{1,0}}$). In normalized shot-noise units, $\hat{X}_{in_{1,0}}:=(1/\sqrt{2})(\hat{X}_{+_{1,0}} + \hat{X}_{-_{1,0}})$, with statistical properties\footnote{Note, throughout this work we let $\hbar=2$. Also, in shot-noise units, $\hat{X}=\hat{a}+\hat{a}^{\dag}$.}
\begin{align}
    \nonumber
    &\langle\hat{X}_{in_{1,0}}\rangle = \frac{\sqrt{2}d}{w_0}\left(\sqrt{N_-}\cos(\Phi_{-})-\sqrt{N_+}\cos(\Phi_{+})\right),~\text{and}\\
    &\langle\delta\hat{X}_{in_{1,0}}^2\rangle = \langle\hat{X}_{in_{1,0}}^2\rangle-\langle\hat{X}_{in_{1,0}}\rangle^2 = 1,
\end{align}
Here, $\hat{X}_{\pm_{1,0}}$ denotes the $\hat{X}$ quadrature of sources $A_{\pm}$, respectively, in the $\text{HG}_{1,0}$ mode, and $\langle\hat{X}_{in_{1,0}}\rangle$ is the mean and $\langle\delta\hat{X}_{in_{1,0}}^2\rangle$ is the variance. We use the property $\langle\hat{a}_{1,\pm}\rangle=\mp d\sqrt{N_{\pm}}\exp(i\Phi_{\pm})/w_0$, and $\langle\hat{a}^\dag_{1,\pm}\rangle=\langle\hat{a}_{1,\pm}\rangle^*$. The quadrature measurement is a Gaussian distribution which can be expressed as $\text{Pr}_{1,0}(\hat{X}_{in_{1,0}}|d)~\sim~\mathcal{N}\left(\langle\hat{X}_{in_{1,0}}\rangle,\langle\delta\hat{X}_{in_{1,0}}^2\rangle\right)$, where $\mathcal{N}(\cdot,\cdot)$ denotes a normal distribution. 
For the purpose of notation, we will refer to $\hat{X}_{in_{1,0}}$ simply as the random variable $X$ below. It is straightforward to now derive the FI of the BHD measurement,
\begin{align}
    \label{eq:FI}
   \nonumber \mathcal{F}_{\text{BHD}} = \int_{-\infty}^{\infty} dX \left(\frac{\partial}{\partial d}\ln\text{Pr}_{1,0}(X|d)\right)^2\text{Pr}_{1,0}(X|d) \\
   = \frac{2}{w_0^2}\left(\sqrt{N_-}\cos\left(\Phi_{-}\right)-\sqrt{N_+}\cos\left(\Phi_{+}\right)\right)^2.
\end{align}
Clearly, the FI evades Rayleigh's curse since in the regime $N_{\pm}\ll1$, $\mathcal{F}_{\text{BHD}}$ remains independent of $d$ \cite{Yang2017a}, and this means the BHD setup supersedes direct imaging (which suffers from the Rayleigh's curse) when $d<w_0$. However, \cref{eq:FI} is phase dependent due to the terms $\Phi_{\pm}$, which is a crucial difference from prior works \cite{Tsang2016}.

The CRB is the mean square error that is approached by maximum likelihood estimation of an infinite number of samples of the quadrature \cite{Petz2011,Liu2020}, and is found by taking the inverse of the FI in \cref{eq:FI}. It follows that the SQL, the \textit{root} mean square error, is then 
\begin{align}
    \label{eq:rmsd}
    \nonumber
    d_{\text{SQL}} &= (\mathcal{F}_{\text{BHD}})^{-{1/2}} \\
    &= \frac{w_0}{\sqrt{2}\left(\sqrt{N_-}\cos\left(\Phi_{-}\right)-\sqrt{N_+}\cos\left(\Phi_{+}\right)\right)}.
\end{align}
Here, $d_{\text{SQL}}\propto1/\sqrt{N_{\pm}}$ typifies the definition of the SQL \cite{Pinel2012b}. Further, we emphasize that our analysis extends the result in \cite{Sun2014} to two sources with independent photon numbers and relative phase terms.

In practice, we want to minimize $d_{\text{SQL}}$ as much as possible in order to reduce the mean square error and maximize the precision of the BHD setup. Evidently, $\Phi_{\pm}$ and $N_{\pm}$ play vital roles in minimizing $d_{\text{SQL}}$. For instance, the denominator is maximized only when $\lvert\Phi_{-} - \Phi_{+}\rvert=n\pi$ $\forall ~n\in\mathrm{Z}$. The relative phase from each source can be defined as $\Phi_{\pm}:=\phi_{\pm}+2\pi\ell_{\pm}/\lambda$, where the former component denotes a time-based offset and the latter denotes a path-based offset. Therefore, to control $\lvert\Phi_{-} - \Phi_{+}\rvert$ during the entire measurement time (over all captured samples of the $\hat{X}_{in,_{1,0}}$ quadrature), we require perfect clock synchronization and path stabilization. For the time being, we will assume $\lvert\Phi_{-} - \Phi_{+}\rvert=n\pi$, emphasizing that our analysis presents a best-case scenario in terms of phase error. In \cref{sec:stellar}, we relax this phase error assumption to evaluate a more general scenario.

Lastly, we note that maximizing $N_{\pm}$ would require capturing more photons from the sources by either using a higher intensity source field, or averaging over more samples for a longer measurement time. In practice, the source intensity will be limited by the size, weight and power (SWaP) constraints typical of a satellite. Further, the high orbital velocity of satellites in LEO would limit the time within which all satellites are appropriately visible for transmission. Therefore, we expect that $N_{\pm}$ is constrained in practice.

\subsubsection{\label{sec:snr}SNR analysis}
The SQL is the ``best-case" (quantum noise-limited) performance for a setup that uses coherent sources with given $w_0$, $N_{\pm}$ and $\Phi_{\pm}$. In a real LEO deployment, however, the actual performance would be considerably worse due to various factors that contribute loss and noise to the statistical properties of the quadrature measurements. 

The performance of the BHD setup in a realistic LEO environment would be firstly impacted by diffraction and photodetection inefficiency. Intuitively, a receiver with a finite aperture would capture a fraction of the total incoming light over long inter-satellite distances (between source and receiver) due to diffraction. In addition, inefficiencies in the BHD photodetectors will reduce the SNR of measurements. Here, we analyze both sources of photonic loss, and seek to find thresholds within which our system may still achieve super-resolution performance.

\begin{figure}
    \centering
    \includegraphics[width=\linewidth]{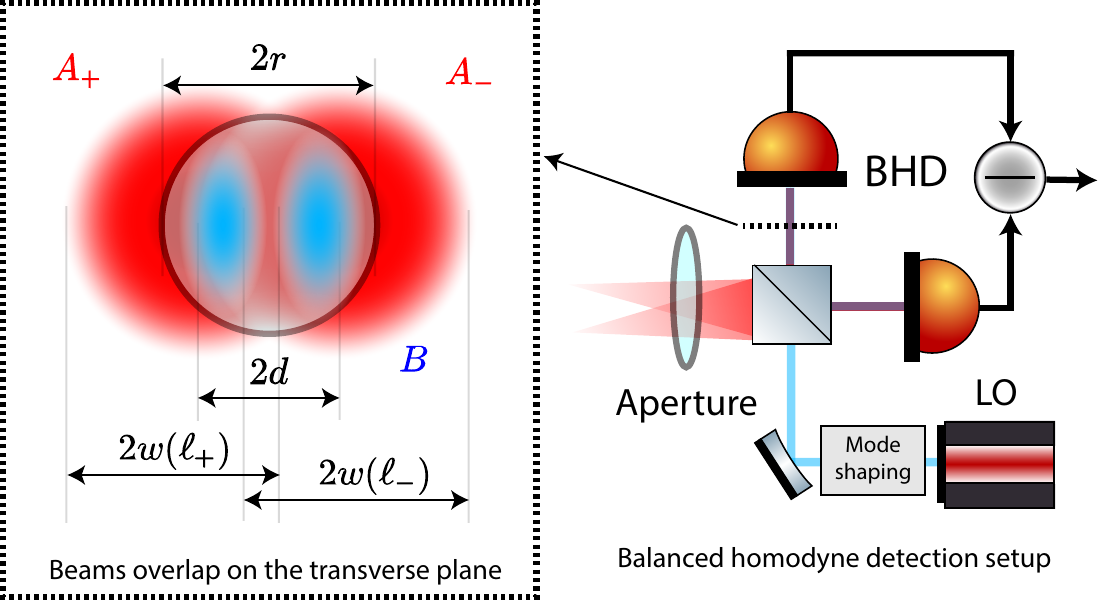}
    \caption{\label{fig:overlap-coefficient}On the left, we show the spatial overlap between the light from $A_+$ and $A_-$ (red) and the light from $B$'s LO (blue). $B$ will measure the portion of the $A_+$ and $A_-$ fields within the aperture $-r\leq x\leq r$. On the right, we show the optical path from the received beams to the BHD output.}
\end{figure}

Diffraction loss on each source can be modeled using a fictitious beam splitter (BS) which transforms the individual fields in \cref{eq:e-in} as follows
\begin{align}
    \label{eq:e-t}
    \nonumber \hat{E}_{\pm}(x,y,\ell_{\pm}) \rightarrow  \sqrt{\eta T(x,y,\ell_{\pm})}\hat{E}_{\pm}(x,y,\ell_{\pm})+\\
    \left(\sqrt{\eta(1-T(x,y,\ell_{\pm}))}+\sqrt{1-\eta}\right)\hat{v}. 
\end{align}
Here, $\hat{v}$ denotes the vacuum mode contribution from the channel, $\eta$ denotes the photodetection efficiency in the BHD ($0\leq\eta\leq1$), and $T(x,y,\ell_{\pm})$ is the channel transmissivity ($0\leq T(x,y,\ell_{\pm}) \leq1$). Recall our assumption that $\theta_d=0$ in \cref{eq:e-in} and that $d$ is entirely along the $x$-axis. In this case, we can remove the $y$-axis degree of freedom, $T(x,y,\ell_{\pm})\rightarrow T(x,\ell_{\pm})$, and evaluate
\begin{align}
    \nonumber T(x,\ell_{\pm}) &= \int_{-r}^{r}dx~u_1(x,\ell_{\pm})u^*_1(x,\ell_{\pm}) \\
    &= \erf\left(\frac{\sqrt{2}r}{w(\ell_{\pm})}\right)-\frac{2^{3/2}r\exp\left(-\frac{2r^2}{w^2(\ell_{\pm})}\right)}{\sqrt{\pi}w(\ell_{\pm})} := T_{\pm}.
\end{align}
Let us henceforth simply denote $T(x,\ell_{\pm})\rightarrow T_{\pm}$. See \cref{fig:overlap-coefficient} for a visualization of the receiver setup. 

The attenuated fields in \cref{eq:e-t} will enter $B$'s aperture and mix with the LO on board during BHD. For ideal projection of the incoming fields onto $\text{HG}_{1,0}$, $B$'s LO needs to be perfectly mode-matched to $u_{1,0}(x,y,\ell_{\pm})$, with positive electric-field operator
\begin{align}
    \hat{E}_{lo}(x,y,\ell_{\pm}) = \mathcal{E}u_{1,0}(x,y,\ell_{\pm})\hat{b}_{1},
\end{align}
where $\hat{b}_1$ denotes the LO annihilation operator in the $\text{HG}_{1,0}$ mode. 

The BHD output signal can be derived using the standard process outlined in our previous works \cite{Gosalia2022,Gosalia2023}. Let us consider each source individually, source $\hat{A}_{\pm}$ produces BHD output $\hat{J}_{\pm}$, given by
\begin{align}   
    \label{eq:j-stat}
    \nonumber \hat{J}_{\pm} = \int_{-r}^{r} dx~\hat{E}_{\pm}^\dag(x,\ell_{\pm})\hat{E}_{lo}(x) + \hat{E}_{\pm}(x,\ell_{\pm})\hat{E}_{lo}^\dag(x), \\
    \nonumber 
    = \mathcal{E}^2 \bigg[\sqrt{\eta T_{\pm}}\left(\hat{a}_{1,\pm}\hat{b}^\dag_1+\hat{a}^\dag_{1,\pm}\hat{b}_1\right)+\\
    \left(\sqrt{\eta(1-T_{\pm})}+\sqrt{1-\eta}\right)\left(\hat{v}\hat{b}_1^\dag+\hat{v}^\dag\hat{b}_1\right)\bigg].
\end{align}
The total BHD output from both sources is simply $\hat{J}_{in}~=~(1/\sqrt{2})(\hat{J}_+~+~\hat{J}_-)$. The statistical properties of $\hat{J}_{in}$ are\footnote{Throughout this work, all variances are taken at zero-mean condition when $d=\bar{d}=0$.} 
\begin{align}
    \nonumber 
    \langle\hat{J}_{in}\rangle = \frac{\mathcal{E}^2\sqrt{2\eta N_{lo}}d}{w_0}\left(\sqrt{T_{+}N_{+}}+\sqrt{T_{-}N_{-}}\right),~\text{and}\\
    \langle\delta\hat{J}_{in}^2\rangle =~\mathcal{E}^4N_{lo}\bigg[1+\sqrt{\eta(1-\eta)}\left(\sqrt{1-T_{+}}+\sqrt{1-T_{-}}\right)\bigg].
\end{align}
Note, here we have used the relations $\langle\hat{b}_{1}\rangle=\sqrt{N_{lo}}\exp(i\Phi_{lo})$, and set $\Phi_{lo}=0=\Phi_{-}$ and $\Phi_{+}=\pi$. 

As per previous studies \cite{Lamine2008,Delaubert2008,Brossel2018}, here we define the SNR as $\mathcal{S}=\langle\hat{J}_{in}\rangle/\sqrt{\langle\delta\hat{J}^2_{in}\rangle}$. When $\mathcal{S}=1$, we derive
\begin{align}
    \label{eq:d}
    \nonumber d_{min} = \frac{w_0}{\sqrt{2\eta}\left(\sqrt{T_{+}N_{+}}+\sqrt{T_{-}N_{-}}\right)}\times\\
    \bigg[1+\sqrt{\eta(1-\eta)}\left(\sqrt{(1-T_{+})}+\sqrt{1-T_{-}}\right)\bigg]^{1/2}.
\end{align}
Comparison of \cref{eq:rmsd} with \cref{eq:d} shows that both are equal in the perfect lossless case when $\eta=1=T(\ell_{\pm})$. 

\begin{figure}
    \centering
    \includegraphics[width=\linewidth,trim = 3cm 9cm 4cm 10cm, clip]{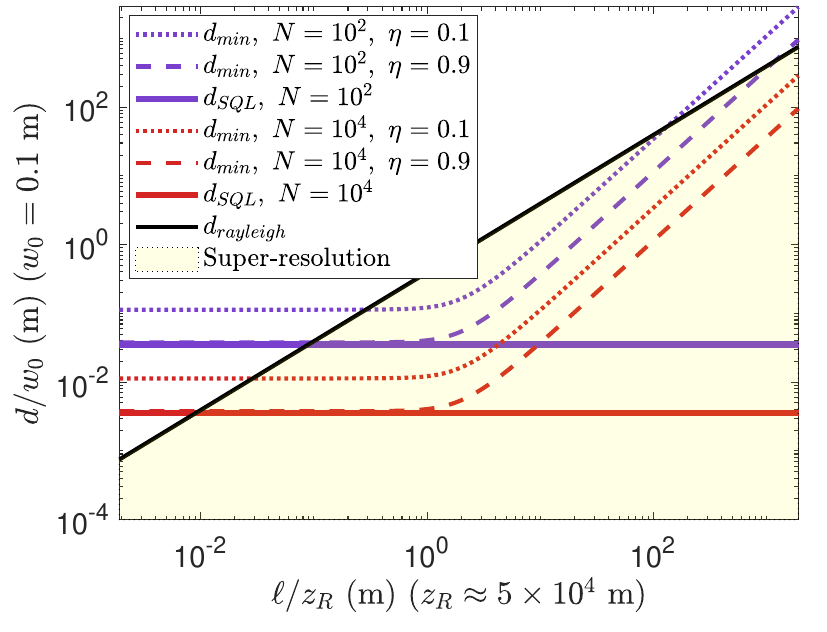}
    \caption{\label{fig:d-example}An example LEO satellite configuration with the system parameters given in \cref{tab:constant-params} undergoing photonic loss due to diffraction and photodetection inefficiency. The yellow region highlights when $d_{min}<d_{rayleigh}$ and thus super-resolution is still achievable. This particular setup exemplifies that BHD with an $\text{HG}_{1,0}$ mode LO can be used to achieve super-resolution over $\ell$ practical of LEO satellites.}
\end{figure}

To visualize the impact of diffraction and detector inefficiency, let us consider an example LEO setup between three satellites ($A_{\pm}$ and $B$) with system parameters per \cref{tab:constant-params}. For simplicity, we assume that $A_{\pm}$ are equally bright, $N_{+}=N_{-}\equiv N$. In \cref{fig:d-example}, we show the results for two example LEO setups capturing $N=10^2$ and $N=10^4$ photon, and, for both setups, three photodetection efficiency cases are considered: the SQL ($\eta=1$), $\eta=0.9$ (realistic) and $\eta=0.1$ (worst-case). We approximated $\ell_{\pm}\approx\ell$, and spanned $\ell$ over realistic LEO inter-satellite ranges (i.e. $10^5\leq\ell\leq10^7$ m) \cite{Shehata2021}. 

Evidently, super-resolution of $d$ (i.e. $d_{min}<d_{rayleigh}$) is achievable under certain conditions. Notably, the results show tolerance for photodetection inefficiency as super-resolution is achieved even when $\eta=0.1$ within the region $10^4<\ell<10^5$ m. Although, improving $\eta>0.1$ clearly increases the range of suitable $\ell$. Lastly, we note that the diffraction-based loss\footnote{Diffraction causes the rapid expansion of the source beam width for $\ell>z_R$ where $z_R\approx5\times10^4$~m in the example LEO setup.} in the range $\ell>10^5$~m, still allows for super-resolution in some cases. This final point indicates the robustness of the BHD solution against photonic loss from diffraction and photodetection inefficiency.

\begin{table}
    \centering
    \caption{\label{tab:constant-params}System parameters used in the example LEO setup.}
    \begin{tabular}{|c|c|c|c|c|c|c|c|}
        \hline
        $\lambda$ & $w_0$ & $r$ & $\Phi_{lo},\Phi_+$ & $\Phi_{-}$ & $N_{lo}$ & $N_{\pm},N/2$\footnote{\label{uos}Unless otherwise specified.} & $\eta$\footref{uos}\\
        \hline
        $600$~nm & $0.1$~m & $0.2$~m & $0$ & $\pi$ & $10^6$ & $10^2$ & $0.9$ \\
        \hline
    \end{tabular}
\end{table}

\subsection{\label{sec:centroid}Centroid misalignment}
Thus far our analysis assumes that $B$ is perfectly aligned with an \emph{a priori} known centroid that coincides with the weighted center of the two sources\footnote{Weighted, for instance, by the relative brightness of each source.}. In practice, however, $B$'s pointing direction may fluctuate about a \textit{true} centroid or contain a fixed offset due to practical satellite pointing challenges such as on-board mechanical vibrations \cite{Vilnrotter,Farid2007,Madni2021}. Fluctuating and fixed centroid misalignments negatively impacts the achievable $d_{min}$ in a BHD setup. Here, we evaluate both sources of misalignment on the ability to achieve super-resolution. 

\subsubsection{Fluctuating misalignment}
\label{sec:centroid-fluctuating}
We first consider the fluctuating misalignment case where $B$ jitters about a true centroid with mean $0$ and standard deviation of $\sigma_d$ along the $x$-axis. Pointing fluctuations at $B$ can be equivalently modeled as a fluctuating measurement of $d$ by setting $d\rightarrow D$, where $D~\sim~\mathcal{N}(\bar{d}, \sigma_d^2)$, with mean $\bar{d}$. We can now re-derive \cref{eq:j-stat} with $d\rightarrow D$ to find\footnote{For a normally distributed variable, $X$, $\mathbb{E}[X]=\mu$ while $\mathbb{E}[X^2]~=~\sigma^2+\mu^2$ where $\mu$ is the mean, $\sigma^2$ is the variance and $\mathbb{E}[\cdot]$ denotes the expectation.}
\begin{align}
    \label{eq:dmin-D-stats}
    \nonumber \langle\hat{J}_{in}\rangle_D = \frac{\mathcal{E}^2\sqrt{2\eta N_{lo}}\bar{d}}{w_0}\left(\sqrt{T_{+}N_{+}}+\sqrt{T_{-}N_{-}}\right), \\
    \langle\delta\hat{J}^2_{in}\rangle_D = \langle\delta\hat{J}^2_{in}\rangle + \frac{\mathcal{E}^4\eta\sigma_d^2}{w_0^2}\left(T_{+}N_++T_{-}N_-\right). 
\end{align}
Evidently, when $\sigma_d\neq0$, $\langle\delta\hat{J}^2_{in}\rangle_D$ increases --- thus, a fluctuating centroid contributes noise to the final BHD measurement. The impact of this noise on $d_{min}$ can be evaluated by following the same SNR analysis as before,
\begin{align}
    \label{eq:dmin-D}
    \nonumber
    \bar{d}_{min,D} = \frac{w_0}{\sqrt{2\eta N_{lo}}\left(\sqrt{T_+N_+}+\sqrt{T_-N_-}\right)}\times\\
    \nonumber
    \bigg[N_{lo}\left(1+\sqrt{\eta(1-\eta)}\left(\sqrt{1-T_{+}}+\sqrt{1-T_{-}}\right)\right)\\
    +\frac{\eta\sigma_d^2}{w_0^2}\left(T_{+}N_++T_{-}N_-\right)\bigg]^{1/2}.
\end{align}

\begin{figure}
    \centering
    \includegraphics[width=\linewidth,trim = 3cm 9cm 4cm 10cm, clip]{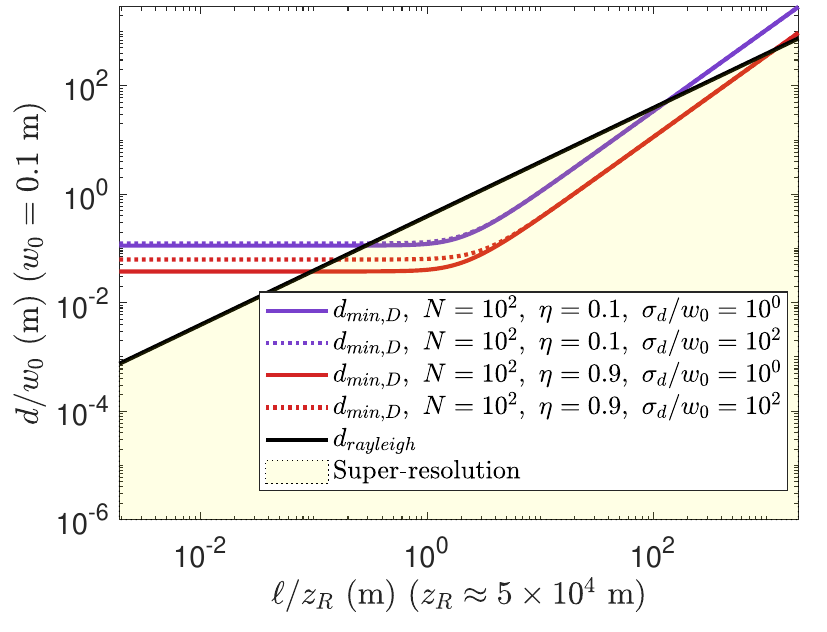}
    \caption{\label{fig:cmis}The example LEO setup repeated with a fluctuating centroid misalignment issue. Certain configurations of $\sigma_d$, $\ell$ and $\eta$ here do fall within the $d_{min}<d_{rayleigh}$ region and thus indicate super-resolution precision in LEO.} 
\end{figure}

In \cref{fig:cmis} we perform the super-resolution check, $d_{min}<d_{rayleigh}$, on the previous example LEO setup with the more realistic $N=10^2$ source photons for various $\eta$ and $\sigma_d$. We considered two cases of $\sigma_d$, including $\sigma_d/w_0=1$ (realistic) and $\sigma_d/w_0=10^3$ (worst-case) \cite{Madni2021} with corresponding best-worst case of $\eta$ to show that super-resolution is achievable over reasonable $\ell$. 

\begin{figure}
    \centering
    \includegraphics[width=\linewidth,trim = 3cm 9cm 4cm 10cm, clip]{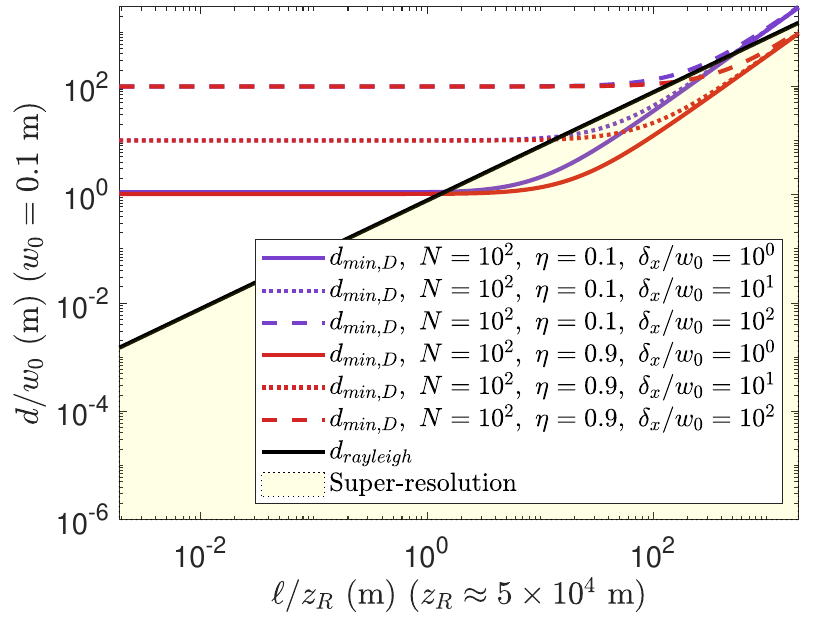}
    \caption{\label{fig:cmis_fixed}The example LEO setup repeated with a fixed centroid misalignment issue. The severity of $\delta_x$ is evident by comparison with \cref{fig:cmis}. Evidently, controlling $\delta_x$ is crucial to maximize the range for super-resolution.}
\end{figure}

\subsubsection{Fixed misalignment}
\label{sec:centroid-fixed}
We now consider a fixed misalignment issue along the $x$-axis, denoted $\delta_x$, relative to the true centroid. The fixed misalignment can be expressed as an additional term in the source spatial-mode profile in \cref{eq:u_0_dx} and expanded using Taylor series to yield
\begin{align}
    \label{eq:u0-deltax}
    u_0\left(x+d-\delta_x,\ell_{\pm}\right) \approx u_0(x,\ell_{\pm})-\frac{d-\delta_x}{w_0}u_1(x,\ell_{\pm}).
\end{align}
Again, we can re-derive \cref{eq:j-stat} with $\delta_x$ to find
\begin{align}
    \nonumber \langle\hat{J}_{in}\rangle_{\delta_x}=\langle\hat{J}_{in}\rangle -\frac{\mathcal{E}^2\sqrt{2\eta N_{lo}}\delta_x}{w_0}\left(\sqrt{T_{+}N_+}+\sqrt{T_{-}N_-}\right),\\
    \text{and}~\langle\delta\hat{J}^2_{in}\rangle_{\delta_x} = \langle\delta\hat{J}^2_{in}\rangle + \frac{\mathcal{E}^4\eta\delta_x^2}{w_0^2}\left(T_{+}N_++T_{-}N_-\right). 
\end{align}
Contrasting with the fluctuating misalignment results in \cref{eq:dmin-D-stats}, $\delta_x$ influences both the mean and the variance --- reducing the signal strength (loss) and increasing the noise of the final BHD measurement. The impact of $\delta_x$ on $d_{min}$ is consequently given by 
\begin{align}
    \label{eq:dmin-deltax}
    \nonumber d_{min,\delta_x}=\frac{w_0}{\sqrt{2\eta N_{lo}}\left(\sqrt{T_+N_+}+\sqrt{T_-N_-}\right)}\times\\
    \nonumber
    \bigg[N_{lo}\left(1+\sqrt{\eta(1-\eta)}\left(\sqrt{1-T_{+}}+\sqrt{1-T_{-}}\right)\right)\\
    +\frac{\eta\delta_x^2}{w_0^2}\left(T_{+}N_++T_{-}N_-\right)\bigg]^{1/2}+\delta_x.
\end{align}
We now have a fixed $\delta_x$ that proportionately offsets the minimum $d_{min,\delta_x}$, correspondingly reducing the precision of our BHD setup. Fixed centroid misalignment is therefore a significant issue for super-resolution especially in BHD setups. 

In \cref{fig:cmis_fixed}, the example LEO setup is repeated with varying levels of $\delta_x$, against the super-resolution check. By comparing \cref{fig:cmis} with \cref{fig:cmis_fixed}, it is evident that $d_{min}$ is more sensitive to a fixed misalignment than a fluctuating misalignment as there is an even smaller region of $\ell$ in which the super-resolution check is satisfied. For any meaningful implementation over LEO satellites, the fixed centroid misalignment issue will need to be reasonably controlled.


We should caution that a key requirement for ``good" BHD performance is that the LO is a good match with the spatial and temporal mode profiles of the source. In our analysis, we have implicitly assumed that this is accounted for by the use of accurate pilot symbols as a means to optimally map the phase correction between the signal and the LO. If such a phase correction was not in place a form of phase-averaged homodyne detection \cite{Shapiro2009} would be required. This would obviously lead to a reduction in the performance we have reported here. Such an effect is most likely more important in the context of incoherent (thermal) sources.


\subsection{\label{sec:stellar}Thermal Sources}
BHD can also be applied to super-resolve the separation between thermal sources of photons such as stellar systems \cite{Zanforlin2022,Tsang2019,Tsang2018,Yang2017,Yang2017a,Tsang2016,Nair2016}.
Taking the lead from \cite{Yang2017a}, we model a stellar source as a thermal state with the $\hat{X}$ quadrature of the $\text{HG}_{1,0}$ mode distributed as $\hat{X}_{in_{1,0},T}\sim \mathcal{N}\left(0,d(N_{+,T}+N_{-,T})/w_0+1\right)$. 
Here $N_{\pm,T}$ are the average photons in the thermal state from each stellar source $A_{\pm,T}$ respectively.
The FI of $d$ between two thermal sources is then expressed as
\begin{align}
    \mathcal{F}_{\text{BHD},T} = \frac{d^2\left(N_{+,T}+N_{-,T}\right)}{\bigg(d^2(N_{+,T}+N_{-,T})+w_0^2\bigg)^2}.
\end{align}
As already evidenced by our analysis, a practical LEO implementation would perform sub-optimally in comparison to the FI due to losses and noise issues from diffraction, photodetection inefficiency, centroid misalignment, and potentially other loss and noise issues. 

For the stellar case, we can use the same analysis presented here up to \cref{eq:j-stat}. However a caveat is that thermal sources do not have the same temporal coherence properties as laser sources nor is there a well defined phase relationship between the signal and LO immediately accessible. In such cases, a time-averaged version of the BHD measurement is likely more useful \cite{Yang2017a,Frank2023}, denoted here as the average power given by \cite{Shapiro2009}
\begin{align}
    \nonumber
    \langle\hat{P}_{in}\rangle = \langle\hat{P}_{+}\rangle+\langle\hat{P}_-\rangle~\text{where}~\hat{P}_{\pm} = \int_{0}^{t_{m}}dt~\hat{J}_{\pm}\hat{J}^\dag_{\pm}.
\end{align}
Here, $\hat{P}_{\pm}$ is the power received at $B$ from each source and $t_{m}$ is the measurement time. We can now use the density operator description of our thermal sources (extending the coherent source description in \cref{eq:coherent-sources})
\begin{align}
    \nonumber
    \rho_{in} = \frac{1}{2}\left(\rho_{+}+\rho_{-}\right),~\text{where}\\
    \nonumber 
    \rho_{\pm} \approx \frac{1}{\pi N_{\pm,T}}\int d^2\alpha~{\rm e}^{-|\alpha|^2/N_{\pm,T}}\times\\
    \ket{\alpha}_{0,0}\bra{\alpha}\otimes\ket{\mp\frac{d}{w_0}\alpha}_{1,0}\bra{\mp\frac{d}{w_0}\alpha},
\end{align}
to find the expectation of $\hat{P}_{in}$. Namely, under ideal conditions (i.e. $T_{\pm}=\eta=1$),
\begin{align}
    \nonumber
    \label{eq:Pin}
    \langle\hat{P}_{in}\rangle = |\mathcal{E}|^4\left[2N_{lo}+\frac{d^2}{w_0^2}(2N_{lo}+1)\left(N_{+,T}+N_{-,T}\right)\right]\\
    \approx2|\mathcal{E}|^4N_{lo}\left[1+\frac{d^2}{w_0^2}\left(N_{+,T}+N_{-,T}\right)\right], 
\end{align}
when $N_{lo}\gg N_{\pm,T}$ --- the usual case for BHD. From \cref{eq:Pin}, it is evident that $\langle\hat{P}_{in}\rangle$ will have a non-zero power level when $d=0$, which will need to be calibrated as the baseline. Consequently, when $d>0$ and $N_{\pm,T}>0$, the measurement will exceed this baseline. Further, for competitive performance that exceeds direct imaging we require that $N_{+,T}+N_{-,T}>2$ \cite{Yang2017a}, although sub-optimal estimation of $d$ is still possible below this limit. 



\section{\label{sec:conc}Conclusion}
A practical implementation of super-resolution on LEO satellites demands a holistic assessment of all performance limiting factors. We presented a study that evaluates the impact of loss and noise effects expected on a real LEO deployment. The loss and noise factors considered included diffraction, photodetection inefficiency, and fluctuating and fixed centroid misalignment. Using an example LEO satellite setup with configuration per \cref{tab:constant-params}, we quantified the scenarios in which super-resolution precision could still be achieved despite these issues. As a key takeaway, we find that a fixed centroid misalignment is a significant issue that needs to be controlled in practice within approximately $\delta_x<w_0$. Comparatively, other factors such as a fluctuating centroid misalignment, photodetection inefficiency and diffraction were less significant. 
Finally, our study can be easily extended to stellar systems, and we discussed some of the subtleties that would be required in this scenario. Evidently, BHD is a promising technique for achieving super-resolution on satellites in LEO, and deserves further investigation. Approved for Public Release: NG23-2418.

\bibliographystyle{ieeetr} 
\bibliography{references1}

\begin{thebibliography}{10}

\bibitem{limitray}
L.~Motka, B.~Stoklasa, M.~D'Angelo, P.~Facchi, A.~Garuccio, Z.~Hradil,
  S.~Pascazio, F.~V. Pepe, Y.~S. Teo, J.~{\v{R}}eh{\'a}{\v{c}}ek, and L.~L.
  S{\'a}nchez-Soto, ``{Optical resolution from Fisher information},'' {\em The
  European Physical Journal Plus}, pp.~130--143, 2016.

\bibitem{Kolobov2000}
M.~I. Kolobov and C.~Fabre, ``{Quantum limits on optical resolution},'' {\em
  Physical Review Letters}, pp.~3789--3792, 2000.

\bibitem{Tamburini2006}
F.~Tamburini, G.~Anzolin, G.~Umbriaco, A.~Bianchini, and C.~Barbieri,
  ``{Overcoming the Rayleigh criterion limit with optical vortices},'' {\em
  Phys. Rev. Lett.}, pp.~1--4, 2006.

\bibitem{Tsang2019}
M.~Tsang, ``{Resolving starlight: a quantum perspective},'' {\em Contemp.
  Phys.}, pp.~279--298, 2019.

\bibitem{Tsang2016}
M.~Tsang, R.~Nair, and X.~M. Lu, ``{Quantum theory of superresolution for two
  incoherent optical point sources},'' {\em Phys. Rev. X}, pp.~1--17, 2016.

\bibitem{Delaubert2008}
V.~Delaubert, N.~Treps, C.~Fabre, H.~A. Bachor, and P.~R{\'{e}}fr{\'{e}}gier,
  ``{Quantum limits in image processing},'' {\em EPL (Europhysics Lett.},
  pp.~1--5, 2008.

\bibitem{Perez-Delgado2012}
C.~A. P{\'{e}}rez-Delgado, M.~E. Pearce, and P.~Kok, ``{Fundamental limits of
  classical and quantum imaging},'' {\em Phys. Rev. Lett.}, pp.~1--5, 2012.

\bibitem{Lupo2020}
C.~Lupo, Z.~Huang, and P.~Kok, ``{Quantum Limits to Incoherent Imaging are
  Achieved by Linear Interferometry},'' {\em Phys. Rev. Lett.}, pp.~1--6, 2020.

\bibitem{Tsang2018}
M.~Tsang, ``{Subdiffraction incoherent optical imaging via spatial-mode
  demultiplexing: Semiclassical treatment},'' {\em Phys. Rev. A}, pp.~1--18,
  2018.

\bibitem{Len2020}
Y.~L. Len, C.~Datta, M.~Parniak, and K.~Banaszek, ``{Resolution limits of
  spatial mode demultiplexing with noisy detection},'' {\em Int. J. Quantum
  Inf.}, pp.~1--16, 2020.

\bibitem{Boucher2020a}
P.~Boucher, C.~Fabre, G.~Labroille, and N.~Treps, ``{Spatial optical mode
  demultiplexing as a practical tool for optimal transverse distance
  estimation},'' {\em Optica}, pp.~1621--1626, 2020.

\bibitem{Tham2017}
W.~K. Tham, H.~Ferretti, and A.~M. Steinberg, ``{Beating Rayleigh's Curse by
  Imaging Using Phase Information},'' {\em Phys. Rev. Lett.}, pp.~1--6, 2017.

\bibitem{Bonsma-fisher2019}
K.~A.~G. Bonsma-Fisher, W.-K. Tham, H.~Ferretti, and A.~M. Steinberg,
  ``Realistic sub-rayleigh imaging with phase-sensitive measurements,'' {\em
  New Journal of Physics}, pp.~1--14, 2019.

\bibitem{Qi2022}
L.~Qi, X.~Tan, L.~Chen, K.~Y. Ng, A.~J. Danner, and M.~Tsang,
  ``{Quantum-Inspired Superresolution for Multiple Incoherent Optical Point
  Sources},'' {\em 2022 Conf. Lasers Electro-Optics, CLEO 2022 - Proc.},
  pp.~3--4, 2022.

\bibitem{Tao2020}
L.~Tao, A.~Green, and P.~Fulda, ``{Higher-order Hermite-Gauss modes as a robust
  flat beam in interferometric gravitational wave detectors},'' {\em Phys. Rev.
  D}, pp.~1--10, 2020.

\bibitem{Zanforlin2022}
U.~Zanforlin, C.~Lupo, P.~W. Connolly, P.~Kok, G.~S. Buller, and Z.~Huang,
  ``{Optical quantum super-resolution imaging and hypothesis testing},'' {\em
  Nature Communications}, pp.~1--9, 2022.

\bibitem{Ram2006}
S.~Ram, E.~S. Ward, and R.~J. Ober, ``{Beyond Rayleigh's criterion: A
  resolution measure with application to single-molecule microscopy},'' {\em
  Proc. Natl. Acad. Sci. U. S. A.}, pp.~4457--4462, 2006.

\bibitem{Deschout2014}
H.~Deschout, F.~C. Zanacchi, M.~Mlodzianoski, A.~Diaspro, J.~Bewersdorf, S.~T.
  Hess, and K.~Braeckmans, ``{Precisely and accurately localizing single
  emitters in fluorescence microscopy},'' {\em Nat. Methods}, pp.~253--266,
  2014.

\bibitem{Chao2016}
J.~Chao, E.~{Sally Ward}, and R.~J. Ober, ``{Fisher information theory for
  parameter estimation in single molecule microscopy: tutorial},'' {\em J. Opt.
  Soc. Am. A}, pp.~B36--B57, 2016.

\bibitem{Hsu2004}
M.~T. Hsu, V.~Delaubert, P.~K. Lam, and W.~P. Bowen, ``{Optimal optical
  measurement of small displacements},'' {\em J. Opt. B Quantum Semiclassical
  Opt.}, pp.~495--501, 2004.

\bibitem{Delaubert2006}
V.~Delaubert, N.~Treps, M.~Lassen, C.~C. Harb, C.~Fabre, P.~K. Lam, and H.-A.
  Bachor, ``{TEM10 homodyne measurement as an optimal small-displacement and
  tilt-measurement scheme},'' {\em Phys. Rev. A}, vol.~74, p.~053823, nov 2006.

\bibitem{Pinel2012b}
O.~Pinel, J.~Fade, D.~Braun, P.~Jian, N.~Treps, and C.~Fabre, ``{Ultimate
  sensitivity of precision measurements with intense Gaussian quantum light: A
  multimodal approach},'' {\em Phys. Rev. A - At. Mol. Opt. Phys.}, pp.~1--4,
  2012.

\bibitem{Yang2017}
F.~Yang, A.~Tashchilina, E.~S. Moiseev, C.~Simon, and A.~I. Lvovsky,
  ``{Far-field linear optical superresolution via heterodyne detection in a
  higher-order local oscillator mode},'' {\em Optica}, pp.~1148--1152, 2016.

\bibitem{Brossel2018}
V.~Delaubert, N.~Treps, M.~Lassen, C.~C. Harb, C.~Fabre, P.~K. Lam, and H.-A.
  Bachor, ``${\text{tem}}_{10}$ homodyne detection as an optimal
  small-displacement and tilt-measurement scheme,'' {\em Phys. Rev. A},
  pp.~1--10, 2006.

\bibitem{Yang2017a}
F.~Yang, R.~Nair, M.~Tsang, C.~Simon, and A.~I. Lvovsky, ``{Fisher information
  for far-field linear optical superresolution via homodyne or heterodyne
  detection in a higher-order local oscillator mode},'' {\em Phys. Rev. A},
  pp.~1--6, 2017.

\bibitem{Kish2020}
S.~P. Kish, E.~Villase{\~{n}}or, R.~Malaney, K.~A. Mudge, and K.~J. Grant,
  ``{Feasibility assessment for practical continuous variable quantum key
  distribution over the satellite‐to‐Earth channel},'' {\em Quantum Eng.},
  pp.~1--16, 2020.

\bibitem{Lamine2008}
B.~Lamine, C.~Fabre, and N.~Treps, ``{Quantum improvement of time transfer
  between remote clocks},'' {\em Phys. Rev. Lett.}, pp.~1--4, 2008.

\bibitem{Gosalia2022}
R.~Gosalia, R.~Malaney, R.~Aguinaldo, J.~Green, and M.~Clampin, ``{Beyond the
  Standard Quantum Limit in the Synchronization of Low-Earth-Orbit
  Satellites},'' {\em 2022 IEEE Latin-American Conf. Commun.}, pp.~1--6, 2022.

\bibitem{Gosalia2023}
R.~Gosalia, R.~Malaney, R.~Aguinaldo, J.~Green, and P.~Brereton, ``{LEO Clock
  Synchronization with Entangled Light},'' 2023.

\bibitem{Tian2019}
Y.~Tian, J.~Zhong, X.~Lin, H.~Yang, and D.~Kang, ``{Inter-Satellite Integrated
  Laser Communication/Ranging Link with Feedback-Homodyne Detection and
  Fractional Symbol Ranging},'' {\em 2019 IEEE Int. Conf. Sp. Opt. Syst. Appl.
  ICSOS 2019}, pp.~1--4, 2019.

\bibitem{Kogelnik1966a}
H.~Kogelnik and T.~Li, ``{Laser Beams and Resonators},'' {\em Proc. IEEE},
  pp.~1312--1329, 1966.

\bibitem{Sun2014}
H.~Sun, K.~Liu, Z.~Liu, P.~Guo, J.~Zhang, and J.~Gao, ``{Small-displacement
  measurements using high-order Hermite-Gauss modes},'' {\em Appl. Phys.
  Lett.}, vol.~104, pp.~1--3, 2014.

\bibitem{Liu2020}
J.~Liu, H.~Yuan, X.~M. Lu, and X.~Wang, ``{Quantum Fisher information matrix
  and multiparameter estimation},'' {\em J. Phys. A Math. Theor.}, pp.~1--71,
  2020.

\bibitem{Li2023}
Z.~Li, Y.~Wang, H.~Sun, K.~Liu, and J.~Gao, ``{Tilt Measurement at the Quantum
  Cramer–Rao Bound Using a Higher-Order Hermite–Gaussian Mode},'' {\em
  Photonics}, pp.~1--8, 2023.

\bibitem{Xia2023}
B.~Xia, J.~Huang, H.~Li, H.~Wang, and G.~Zeng, ``{Toward incompatible quantum
  limits on multiparameter estimation},'' {\em Nat. Commun.}, pp.~1--12, 2023.

\bibitem{Petz2011}
D.~Petz and C.~Ghinea, ``{Introduction to quantum Fisher information},'' in
  {\em Quantum Probab. Relat. Top.}, pp.~261--281, World Scientific, 2011.

\bibitem{Shehata2021}
M.~I. Shehata, Z.~M. Asmaa, H.~A. Fayed, A.~A. {El Aziz}, and M.~H. Aly,
  ``{Optical Inter-Satellite Link over Low Earth Orbit: Enhanced
  Performance},'' {\em J. Phys. Conf. Ser.}, pp.~1--9, 2021.

\bibitem{Vilnrotter}
V.~Vilnrotter, ``{The effects of pointing errors on the performance of optical
  communications systems},'' 1981.

\bibitem{Farid2007}
A.~A. Farid and S.~Hranilovic, ``{Outage capacity optimization for free-space
  optical links with pointing errors},'' {\em J. Light. Technol.},
  pp.~1702--1710, 2007.

\bibitem{Madni2021}
A.~Madni, N.~Bradley, D.~Cervantes, D.~Eldred, D.~Oh, D.~Mathews, and P.~C.
  Lai, ``{Pointing Error Budget Development and Methodology on the Psyche
  Project},'' {\em IEEE Aerosp. Conf. Proc.}, pp.~1--18, 2021.

\bibitem{Shapiro2009}
J.~H. Shapiro, ``{The quantum theory of optical communications},'' {\em IEEE J.
  Sel. Top. Quantum Electron.}, vol.~15, no.~6, pp.~1547--1569, 2009.

\bibitem{Nair2016}
R.~Nair and M.~Tsang, ``{Far-Field Superresolution of Thermal Electromagnetic
  Sources at the Quantum Limit},'' {\em Phys. Rev. Lett.}, pp.~1--5, 2016.

\bibitem{Frank2023}
J.~Frank, A.~Duplinskiy, K.~Bearne, and A.~I. Lvovsky, ``{Passive
  superresolution imaging of incoherent objects},'' {\em Optica},
  pp.~1147--1152, 2023.

\end{thebibliography}

\end{document}